\newcommand{\PaperTitle}
{MINT: Modeling GenAI Impact on Network Traffic}
\newcommand{\PaperDisplayTitle}
{MINT: Modeling GenAI Impact on Network Traffic}

\documentclass[10pt,sigconf]{acmart}
\newcommand{\PaperNumber}{XXX}
\newcommand{\Andrew}[1]{\textcolor{red}{#1}}


\author{%
Andrew Nguyen\textsuperscript{1},
Samson Kempiak\textsuperscript{1},
Agrim Gupta\textsuperscript{2},
Koushik Kar\textsuperscript{1},
Ish Kumar Jain\textsuperscript{1}
}

\affiliation{%
  \institution{%
  \textsuperscript{1}Rensselaer Polytechnic Institute, Troy, NY, USA, \textsuperscript{2}Samsung Research America, Plano, TX, USA
  }
  \country{}
}

\usepackage{svg}
 \usepackage{color}
 \usepackage{graphicx}
 \usepackage[labelformat=simple]{subcaption}
 \usepackage{xspace}
 \usepackage{multirow}
 \usepackage[ruled,vlined]{algorithm2e}
 \usepackage{ulem}
\usepackage{tabularx}
\usepackage{booktabs}
\usepackage{array}
\usepackage{tikz}
\usetikzlibrary{positioning,calc,decorations.pathreplacing}
\usepackage{adjustbox}

\usepackage{amssymb}
\usepackage{pifont}
\usepackage{enumitem}

\usepackage{placeins}

\begin{document}

\acmYear{2026}\copyrightyear{2026}
\acmConference[WiNTECH '26]{20th ACM Workshop on Wireless Network Testbeds, Experimental evaluation \& Characterization}{October 26--30, 2026}{Austin, TX, USA}
\acmBooktitle{20th ACM Workshop on Wireless Network Testbeds, Experimental evaluation \& Characterization (WiNTECH '26), October 26--30, 2026, Austin, TX, USA}
\acmDOI{10.1145/3831662.3844175}
\acmISBN{979-8-4007-2879-2/26/10}

\title[\PaperTitle]{\PaperDisplayTitle}

\begin{abstract}
Generative AI (GenAI) is becoming a mainstream network workload, yet packet-level simulators lack measure\-ment-driven GenAI traffic models. Currently researchers must approximate GenAI services using traditional sources such as file transfer and video streaming, limiting realistic network evaluation of scheduling and capacity planning. We present MINT, a measurement and modeling framework for GenAI network traffic. Using an isolated net\-work-namespace capture pipeline, we collect client-side traces from three LLM providers across four modalities,  cloud and edge servers, and wired and wireless network access points. We find that GenAI modalities exhibit distinct upload/download asymmetry and burst structures that differ from traditional applications. MINT clusters and models these burst regimes and validate empirical burst timing distribution behavior in ns-3 with normalized Wasserstein distances of 2--25\%. Our results also reveal that realistic packet bursts have significantly more variability than constant token generator models. MINT open-sources the first measurement-driven GenAI traffic model for packet-level network simulation.

\end{abstract}

\maketitle

\flushbottom

\section{Introduction}


Generative AI (GenAI) is emerging as a mainstream network workload, with adoption exceeding 50\% in measured markets and applications spanning chatbots, code generation, and image generation~\cite{jin2025end, stanford2026AI, feng2026beamformer}. Unlike video streaming, which typically follows predictable download-heavy ON--OFF buffer-refill patterns, GenAI traffic depends on the prompt modality, server processing time, and response-generation behavior. These factors can produce highly variable upload/download asymmetry and packet-burst timing, which complicates network scheduling and capacity planning. Yet current network studies largely characterize GenAI traffic using coarse aggregate trends~\cite{
cisco2018vni}, rather than packet-level models that capture this variability. Accurately measuring and modeling these dynamics is therefore essential for realistic evaluation of networks carrying GenAI workloads.

\begin{figure}[tb]
    \centering
    \includegraphics[width=0.9\linewidth]{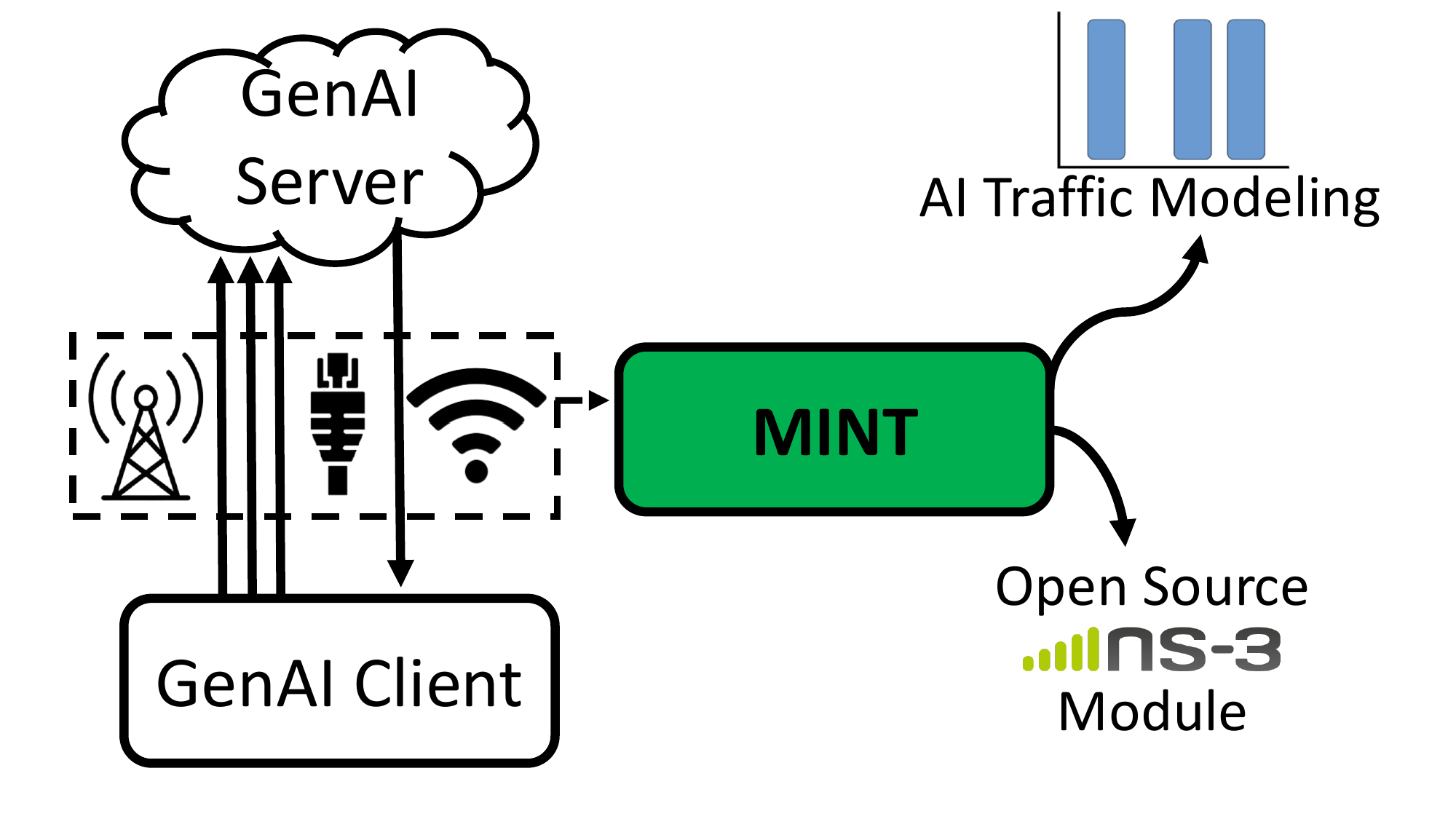}
    \caption{Modeling GenAI Impact on Network Traffic (MINT) with measurements, models, and simulations.
    }
    \label{fig:high-level-diagram}
\end{figure}

We ask two questions: How does GenAI traffic differ from traditional application traffic, such as video streaming, and how can we reproduce its behavior in a packet-level network simulator? Such a model would enable realistic evaluation of scheduling, congestion control, and capacity-planning mechanisms~\cite{sivaroopan2025comprehensive, zink2009characteristics, loh2022youtube}.

Building a measurement-driven GenAI traffic model is challenging for three reasons. \textit{First,} token streaming may not be visible in network traces because the payloads are TLS-encrypted, and providers do not expose how tokens are batched into packets. A machine-in-the-middle (MITM) proxy can recover the text stream over time, but not how it appears as packets on the network. \textit{Second,} client-side captures can be used to measure packet structure, however, they are easily contaminated by browser or background traffic or by transport artifacts such as NIC offloading, which distorts packet sizes and interarrival times. \textit{Finally,} application burst timing must be separated from the access network, since Wi-Fi and 5G add delays that mask application behavior. Therefore, separating the GenAI application traffic from network artifacts is challenging. 

Prior work does not completely address these problems. \textit{GenAI-centric} studies measure server-side token metrics (time-to-first-token, inter-token latency)~\cite{wang2024revisiting, xiao2025streaming} that do not translate into client-side packet bursts, while \textit{client-side} studies pursue other goals, such as token-streaming design or traffic classification, with a single access network, few modalities or prompts, and summary statistics that cannot regenerate packet-level dynamics~\cite{li2024eloquent, tagami2026understanding, alhazbi2025llms}. To the best of our knowledge, no existing GenAI measurement system developed a packet-level simulator with realistic GenAI traffic.

\textbf{MINT}. We present MINT, a measurement and modeling framework for GenAI network traffic, shown in Figure~\ref{fig:high-level-diagram}. MINT isolates direct GenAI API traffic in network namespaces and collects a diverse dataset of nearly 10,000 client-side session traces spanning four modalities, three providers, cloud and edge servers, and Ethernet, Wi-Fi, and 5G access networks. From these measurements, we characterize the upload/download asymmetry and burst structure of GenAI traffic and show how it differs from traditional traffic. We then model the burst behavior at different timescales and implement this model in the open-source Network Simulator 3 (ns-3)~\cite{maza2016framework} and validated ns-3 results with real-world measurements.

MINT addresses four key challenges to build this system. \textit{First,} we identified the stages of a GenAI session using \textit{unencrypted} traffic from an edge server. \textit{Second,} we revealed a multi-timescale burst structure in packet interarrival times---packet serialization, sub-bursts, bursts, and initial server delay---whose timescales can be cleanly separated and modeled independently. \textit{Third,} to isolate GenAI traffic from other unintended traffics, we eliminated contamination at the source by capturing traffic inside a dedicated network namespace that carries only GenAI traffic. \textit{Fourth,} we found that the burst structure is a property of the application rather than the access network. Burst interarrival times are consistent across Ethernet, Wi-Fi, and 5G, so a model can be derived from any one access technology.

\noindent\textbf{Contributions:}
\begin{itemize}[leftmargin=*, nosep]
  \item A diverse, isolated experimental setup spanning Ethernet, Wi-Fi, and a 5G modem, capturing client-side GenAI network traffic across three providers, four modalities, and cloud and edge servers, released as an open trace dataset\footnote{\label{fn:data}\url{https://huggingface.co/datasets/wayslab/llm-network-study-data}} and measurement framework\footnote{\label{fn:framework}\url{https://github.com/wayslab/LLM-Network-Study}}.

  \item A study of GenAI modalities revealing distinct burst timing
  structures and a dynamic UL/DL asymmetry: whereas macro-level reports
  cite a 74\%/26\% downlink/uplink split~\cite{ericsson2025genai}, we
  show that the split depends heavily on modality and application phase, ranging from 0\% to 100\% downlink.

  \item The first open-source, measurement-driven GenAI network burst model in ns-3\footnote{\label{fn:ns3}\url{https://github.com/wayslab/ns-3-dev}}, with accuracy validated based on difference in burst interarrival time distribution of 2-25\% normalized Wasserstein distance. 
\end{itemize}

\section{Background and Related Work}
\subsection{Background}
We review how LLMs generate traffic, how it is measured, and how application traffic is modeled.

\textbf{How do LLM clients and servers interact?} A user prompt is sent to the LLM server, tokenized, and processed in parallel during a prefill phase; the model then generates output tokens iteratively until an end-of-sequence token. Tokens are either streamed to the client in chunks (streaming mode) or returned all at once (non-streaming mode).

\textbf{How is GenAI network traffic measured?}
Tools such as \texttt{tcpdump} and \texttt{Wireshark/tshark} capture packets at points such as Ethernet and Wi-Fi interfaces~\cite{alhazbi2025llms, li2024eloquent}; because payloads are encrypted, analysis relies on metadata such as per-packet size and interarrival time. Traffic is isolated either by attributing packets by IP address and metadata or through controlled experiments with network namespaces, then processed to expose transport-agnostic application behavior~\cite{paxson2002wide,willinger2019lessons, zhang2024quic}.

\textbf{How are network bursts modeled?}
GenAI has no application layer burst traffic model today; a good model must capture the mechanisms that generate the application's traffic. Video streaming follows an ON--OFF buffer-refill model governed by bit rate and buffer state~\cite{bojovic2022enabling, maza2016framework}; file transfer is bulk delivery at link capacity~\cite{wong2006comments}; aggregate models such as Poisson Pareto burst processes reproduce composite traffic but discard per-application structure~\cite{sivaroopan2025comprehensive}. By analogy, streamed GenAI calls for server delay plus token-stream bursts, and non-streamed GenAI for server delay plus a file-transfer-like bulk phase.

\vspace{-.5cm}
\subsection{Related Work}

We group prior work into GenAI-centric measurements and network measurements of GenAI traffic, and show that neither yields a packet-level GenAI traffic model.

\textbf{GenAI-centric measurements.}
A first line of work measures token-level behavior generated at the GenAI server, characterizing application-layer generation dynamics~\cite{qu2025tokenflow, alhazbi2025llms} or designing improved token-streaming algorithms~\cite{li2024eloquent}. These studies establish how tokens are generated and delivered at the application layer. However, token-level metrics do not capture how a provider converts tokens into the packet bursts a client actually receives, so they cannot drive a packet-level network model.


\textbf{Network measurements of GenAI traffic.}
A second line of work measures GenAI traffic on the network itself, building measurement systems for goals such as token streaming over lossy networks, GenAI traffic classification, and high-level impact studies~\cite{koneva2025introducing, cheng2025hello, li2024eloquent, alhazbi2025llms, tagami2026understanding, montieri2026prompts}. These efforts include the first multi-modality measurements of GenAI traffic~\cite{tagami2026understanding} and the first measurements across multiple access networks~\cite{li2024eloquent}, both of which we build on. For traffic modeling, however, three gaps remain. First, no existing measurement system combines the prompt diversity, modality diversity, and access-network isolation that a multi-application burst model requires. Second, prior works characterize burstiness using aggregate statistics~\cite{tagami2026understanding, alhazbi2025llms}, such as coefficient of variation, reducing complex traffic behavior to a small set of summary metrics to compare different GenAI and traditional traffic. Third, existing traffic generators either lack timing patterns or rely on simple models, rather than measurement-driven burst timings and durations. No measurement-driven model has yet reproduced the time-varying packet bursts, received by the client.

\section{MINT Design}
\begin{figure}[!tb]
    \centering
    \includegraphics[width=0.9\linewidth]{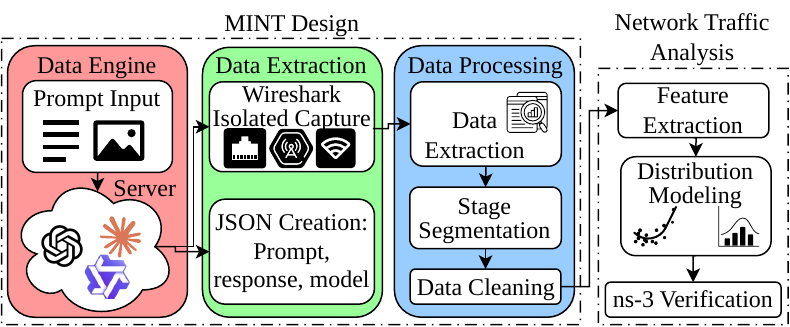}
    \caption{
 End-to-end GenAI traffic measurement processing, modeling, and simulation pipeline.
    }
    \label{fig:testbed}
\end{figure}
MINT has three components, shown in Figure~\ref{fig:testbed}: a \textbf{Data Engine} that replays real-user prompts against cloud and edge LLMs across modalities, a \textbf{Data Extraction} tool that captures each session's traffic in isolation, and a \textbf{Data Processing} stage that segments and cleans traces for modeling. For each prompt, the client starts a capture, sends the prompt to the designated LLM server over one of three access networks---Ethernet, Wi-Fi, or a 5G modem---and stops the capture when the response completes. The \textbf{Network Traffic Analysis} then extracts relevant features and models burst behavior for a network simulator, ns-3, implementation. 

\begin{figure}[!tb]
    \centering
    \includegraphics[width=0.75\linewidth]{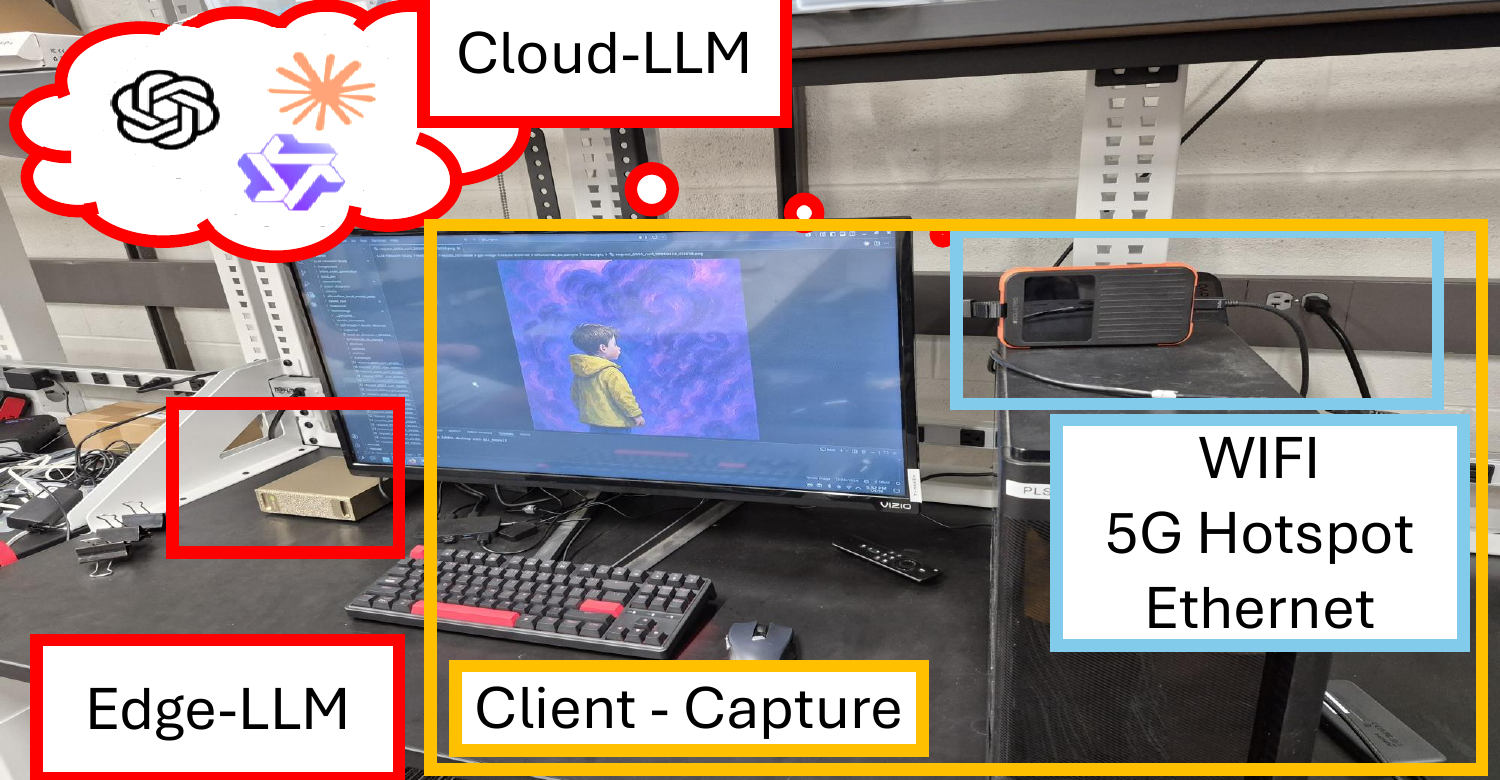}
    \caption{
    Hardware across access types and models.
    }
    \label{fig:hardware_setup}
\end{figure}

\textbf{Hardware Setup}. The experimental setup is shown in Figure~\ref{fig:hardware_setup}. MINT measures proprietary cloud-hosted LLMs, uses an edge-hosted LLM to validate the client-side measurements against ground truth, and repeats captures over wired Ethernet, Wi-Fi, and a SIMO 5G hotspot to validate that our application-layer findings are access-independent.

\subsection{MINT Data Engine}
The MINT data engine automates prompt testing across the model providers in Table~\ref{tab:model_summary} and the modalities in Table~\ref{tab:dataset_summary}, drawing prompts from real-world user prompt datasets.

\textbf{Large Language Models}. We measure three major LLM providers: OpenAI, Anthropic, and open-source Qwen~\cite{yang2025qwen3}, with specific models outlined in Table~\ref{tab:model_summary}. Whereas prior work measured through browsers or apps, we remove that abstraction and access each provider's API directly from Python, eliminating browser and background traffic at the source. We further host an LLM on an edge server to validate our measurements without network artifacts. We measure all three providers to establish that their traffic behavior is repeatable with minor discrepancies. We then focus our modeling on ChatGPT and leave the remaining providers to future work.

\begin{table}[tbp]
\centering
\caption{
Measured models with capture date (MM/DD, 2026), access path, and modality: text-to-text (T2T), image-to-text (I2T), text-to-image (T2I), and code gen.
}
\label{tab:model_summary}
\small
\begin{tabularx}{\columnwidth}{@{}X l l c@{}}
\toprule
\textbf{Model} & \textbf{Date} & \textbf{Access} & \textbf{Modality} \\
\midrule
ChatGPT 5.4  & 07/03 & Browser & T2T  \\
ChatGPT 5.4  & 04/26 & Cloud API & T2T \& I2T \\
ChatGPT Image 1 & 04/28 & Cloud API & T2I \\
Claude Opus 4.6 & 04/25 & Cloud API & T2T \& I2T \\
Qwen 3.5 122B A10B & 06/16 & Cloud API & T2T \& I2T \\
Z-Image-Turbo & 06/16 & Cloud API & T2I \\
Qwen 3.5 122B A10B Int4 & 06/11 & Edge vLLM & T2T \& I2T \\
Z-Image-Turbo & 06/15 & Edge vLLM & T2I \\
Claude Opus 4.8 & 06/29 & Claude Code & Code Gen. \\
ChatGPT 5.5 & 07/03 & Codex & Code Gen. \\
\bottomrule
\end{tabularx}
\end{table}

\textbf{Prompt Datasets \& Inputs}.
Table~\ref{tab:dataset_summary} summarizes the prompt datasets, chosen to cover the modalities that dominate real GenAI usage \cite{stanford2026AI}—text-to-text (T2T), text-to-image (T2I), image-to-text (I2T), and code generation—with 12 short- and long-form YouTube videos for traditional-streaming comparison. These reflect real-user prompts in a reproducible framework: T2T with Berkeley Function Calling Leaderboard (BFCL), T2I and I2T with DiffusionDB image generation prompts. While we do not focus on analyzing the GenAI result, we had to filter the image generation prompts to avoid policy-related restrictions that stopped network traffic abruptly, reducing 400 prompts down to 229. We reused the generated images as image-to-text input. Text and image made up straightforward single-session prompts, so we measured a complex code generation example to show a multi-turn, back-and-forth agentic workflow spawning from one human prompt: 24 agentic prompts over 900 seconds. While these datasets do not reproduce human think-time delays, they represent realistic one-shot and multi-turn prompts reproducibly. Modeling multi-turn prompting is out of scope for this work.

\begin{table}[tbp]
\centering
\caption{
Prompt dataset summary: text-to-text, text-to-image, image-to-text, and code generation (qualitative comparison only*).
}
\label{tab:dataset_summary}
\small
\begin{tabularx}{\columnwidth}{@{}l l l X@{}}
\toprule
\textbf{Source} & \textbf{Name} & \textbf{Subcategory} & \textbf{\# Prompts / Duration} \\
\midrule
\cite{patil2025bfcl} & BFCL & Text-to-text  & 1677 \\
\cite{wang2023diffusiondb} & DiffusionDB & Text-to-Image & 229 \\
\cite{wang2023diffusiondb} & DiffusionDB & Image-to-Text & 229\\
\cite{nyc-taxi-trip-duration} & Kaggle & Code Gen.* & One, 900 seconds \\
\bottomrule
\end{tabularx}
\end{table}

\subsection{MINT Data Extraction Tool}

  Our framework uses a Python client that invokes proprietary and open-source
  GenAI APIs, captures traffic in \texttt{PCAPNG} format, and logs prompt,
  response, and model metadata as \texttt{JSON}; we record average round-trip
  time before and after each run to confirm connection stability. We extract
  traffic three ways depending on cloud, browser, and edge server access. For Cloud APIs for ChatGPT, Claude, and Qwen, we route the client
  through a dedicated Linux network namespace joined to the host by a virtual
  Ethernet (\texttt{veth}) pair. Wireshark captures at the \texttt{veth} interface, isolating traffic from the GenAI flow while retaining LAN access via a consistent local IP. For consumer-facing apps used for comparison, such as ChatGPT Browser and YouTube, we capture from a browser client using Selenium and dedicated Chrome profiles to suppress ads and unrelated browser activity. For the edge server, we host an
  open-source LLM on an NVIDIA DGX Spark with a capture daemon the client
  starts/stops remotely, yielding synchronized client- and server-side traces.
  All three paths produce the same \texttt{PCAPNG} traces and \texttt{JSON}
  metadata.

\textbf{API versus Browser:}  Prior studies measured GenAI traffic through browser/app interfaces, while API calls provide cleaner network isolation and directly observable client–server behavior. Both reflect realistic usage: APIs increasingly support application backends \cite{openai2025enterpriseai}, while browser/app traffic captures direct consumer use. In our comparison, shown in the Appendix, API traffic used a single provider connection with clearly separated interaction stages, whereas browser traffic involved multiple concurrent flows and additional exchanges for the same prompt. These structural differences motivate future decrypted-traffic studies, such as MITM studies, to attribute and model browser GenAI behavior. We focus on MINT API-based traffic.

\subsection{MINT Data Processing Stages}
\label{sec:mint-processing}
After each capture, MINT processes the trace in three steps: data extraction, stage segmentation, and data cleaning.

 \textbf{Data Extraction}.
  From each \texttt{PCAPNG} capture, we extract per-packet size, interarrival time
  (IAT), and direction: with the client fixed at \texttt{10.0.0.2}, packets are
  upload (UL) or download (DL) when the client IP is the source or destination,
  respectively. These fields drive our burst analysis.

\textbf{Stage Segmentation}. Using unencrypted traffic from a local edge server, we identified four stages in a one-shot conversation: handshake, prompt, processing, and response. The handshake begins with a \texttt{Client Hello} and concludes with an ACK packet. The prompt stage follows as sequences of uploaded application data from the client, concluding with an ACK packet from the server. The processing stage is a long period of almost no throughput, measured as the time-to-first-response (TTFR) packet. The response stage then starts with the first server download packet and continues for the rest of the capture. We observed the same structure in encrypted TLSv1.3 API traffic from cloud servers.

\textbf{Data Cleaning}.
We cleaned transport-layer artifacts that impacted TTFR and IAT measurements to ensure we properly model application-layer behavior. TTFR cannot be measured from the first download packet alone because many traces begin with a small 24-byte control packet before the actual response. This would underestimate the true response time. We therefore ignore packets below 50 bytes and measure TTFR from the first \texttt{Application Data} packet.
For IAT measurements, we found 11.5\% of client-side packets were distorted, including oversized superpackets and unusually long packet IATs. Disabling NIC offloading removed the oversized packets, but the timing anomalies remained. We therefore detect these anomalous IATs and replace them using neighboring packet-serialization-scale timing, preventing capture artifacts from contaminating burst-level IAT measurements.

\textbf{Provider Comparison and Repeatability.} These metrics show that measurements are repeatable within a model and similar across providers, supporting our choice to focus modeling on ChatGPT. In multiple trials of each model, the cumulative distribution functions (CDFs) of response packet size and packet IAT overlapped, indicating repeatable experiments, shown in Appendix. Between providers, non-streaming mode showed similar response packet sizes and packet IATs, while streaming mode varied slightly. For example, Qwen produced a longer TTFR than ChatGPT, with similar mean packet IAT but smaller variance.
\section{MINT Network Traffic Analysis}
Using the cleaned packet traces, we answer three questions. First, how does upload (UL) / download (DL) asymmetry change with GenAI applications? Second, how do packet bursts change with GenAI? And third, how can we model and simulate GenAI network traffic?

\subsection{Upload/Download Asymmetry}
GenAI exchanges involve diversity in both uploaded (UL) and downloaded (DL) modality, producing modality-dependent UL/DL behavior. Typically, the fraction of DL is measured across total load, producing one aggregate number such as 74\%/26\% downlink/uplink \cite{ericsson2025genai}. We show the fraction goes through different stages of UL-heavy and DL-heavy, with different throughput and frequency at each stage.

\begin{figure}[!tb]
    \centering
    \includegraphics[width=0.9\linewidth]{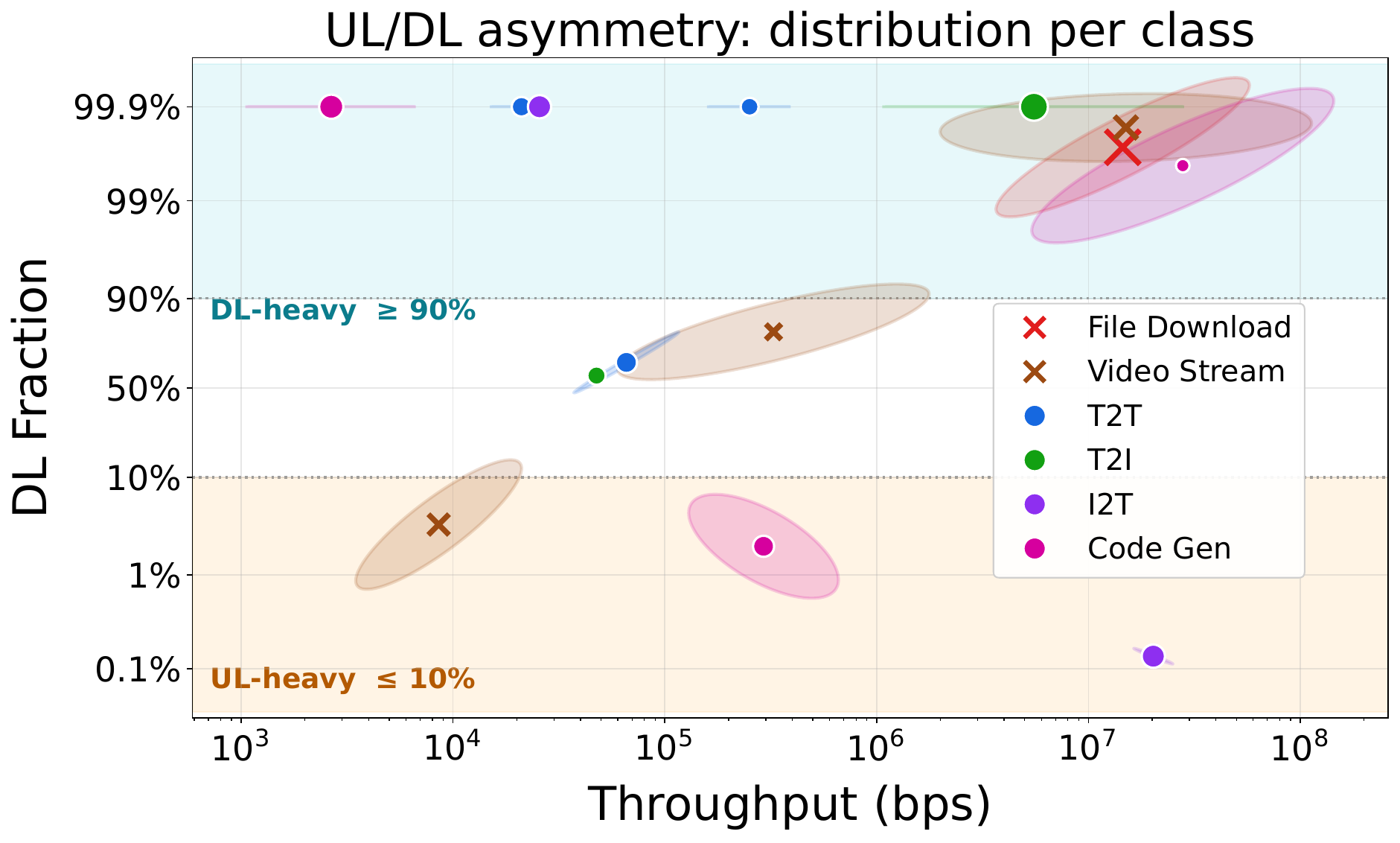}
    \caption{ Gaussian mixture clusters of 1s windows for Download (DL) proportion (y-axis) highlighting changes in DL-heaviness relative to total throughput (x-axis) and packet occurrence (dot size). 
    }
    \label{fig:baseline_vs_genai_1s_windows}
\end{figure}

\textbf{Feature Extraction}.
Since aggregate UL/DL ratios do not highlight the UL/DL asymmetry throughout various stages of traffic, we cluster DL fraction and throughput to show 2 to 3 distinct stages for each application. Across 1-second windows, we compute the total UL and DL bytes, used to compute the ratio of DL to total traffic and the total throughput in bytes per second. We present on a logit scale to highlight the tails of DL-heavy and UL-heavy periods. We cluster data points into 2 to 3 clusters to represent different stages. The size of each dot represents how many windows fall in the cluster. 

\textbf{Analysis}. GenAI has modality-dependent UL/DL stages flowing through UL-heavy and DL-heavy stages, adding specific context to the prior single 74/26 aggregate metric, shown in Figure~\ref{fig:baseline_vs_genai_1s_windows}. 

Traditional application behavior is as expected. File download transfers data very quickly with very little UL occurring. Video streaming has its two largest clusters at DL-heavy and UL-heavy. During DL-heavy, the buffer is filled as a high-throughput download stage. The UL-heavy cluster has very low throughput, representing the periods between buffer refills, with periodic QUIC exchange for YouTube's playback heartbeat. The mixed usage only appears due to window sampling boundaries. 

On the other hand, GenAI sessions show diverse UL/DL asymmetry when comparing stages and modality. Text modalities transfer far less data than image modalities. The T2T scenario highlights the handshake+prompt stages, resulting in moderate throughput, mixed usage, followed by comparably moderate DL-heavy throughput responses. The two DL-heavy clusters represent non-streamed and streamed responses. T2I shows a similar prompt stage, but a much higher DL-heavy throughput in the response stage. Conversely, I2T shows the image upload, much larger than the handshake, with small text responses. Interestingly, the more complex code generation spawns multiple prompt--response exchanges. The prompt stage had moderate throughput, comparable to the text uploads seen in T2T and T2I. However, outside of a single large file download by the agent, the response stage was 100x lower throughput, even lower than the T2T streaming scenarios. Overall, GenAI prompts make traffic more UL-heavy, its DL-heavy responses are smaller than traditional downloads, and both vary greatly with modality.

\subsection{Time-scale Burst Modeling Analysis}

\begin{figure}[!tb]
    \centering

    \begin{minipage}[t]{0.45\linewidth}
        \centering
        \includegraphics[width=\linewidth]{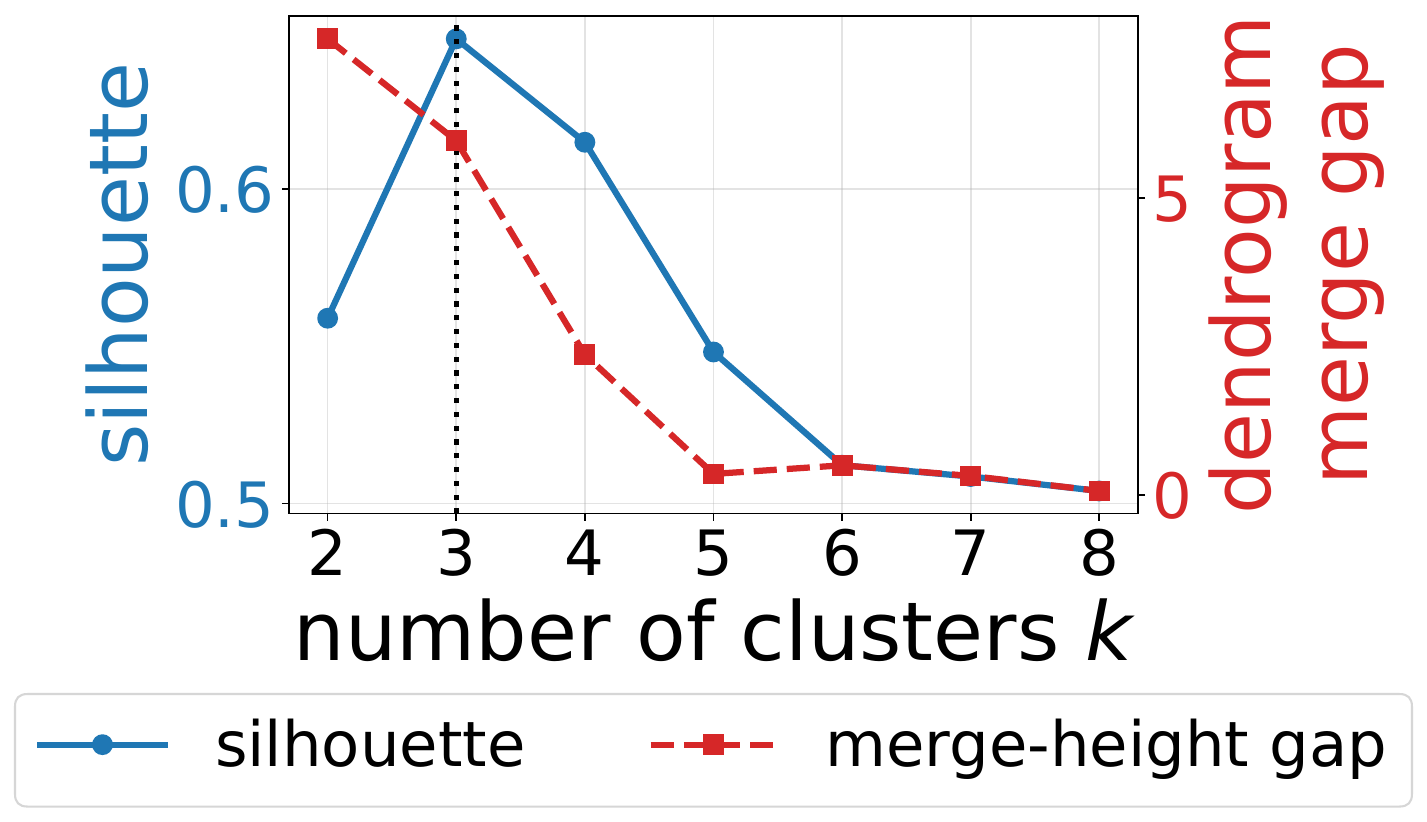}\\
    \end{minipage}
    \hfill
    \begin{minipage}[t]{0.4\linewidth}
        \centering
        \includegraphics[width=\linewidth]{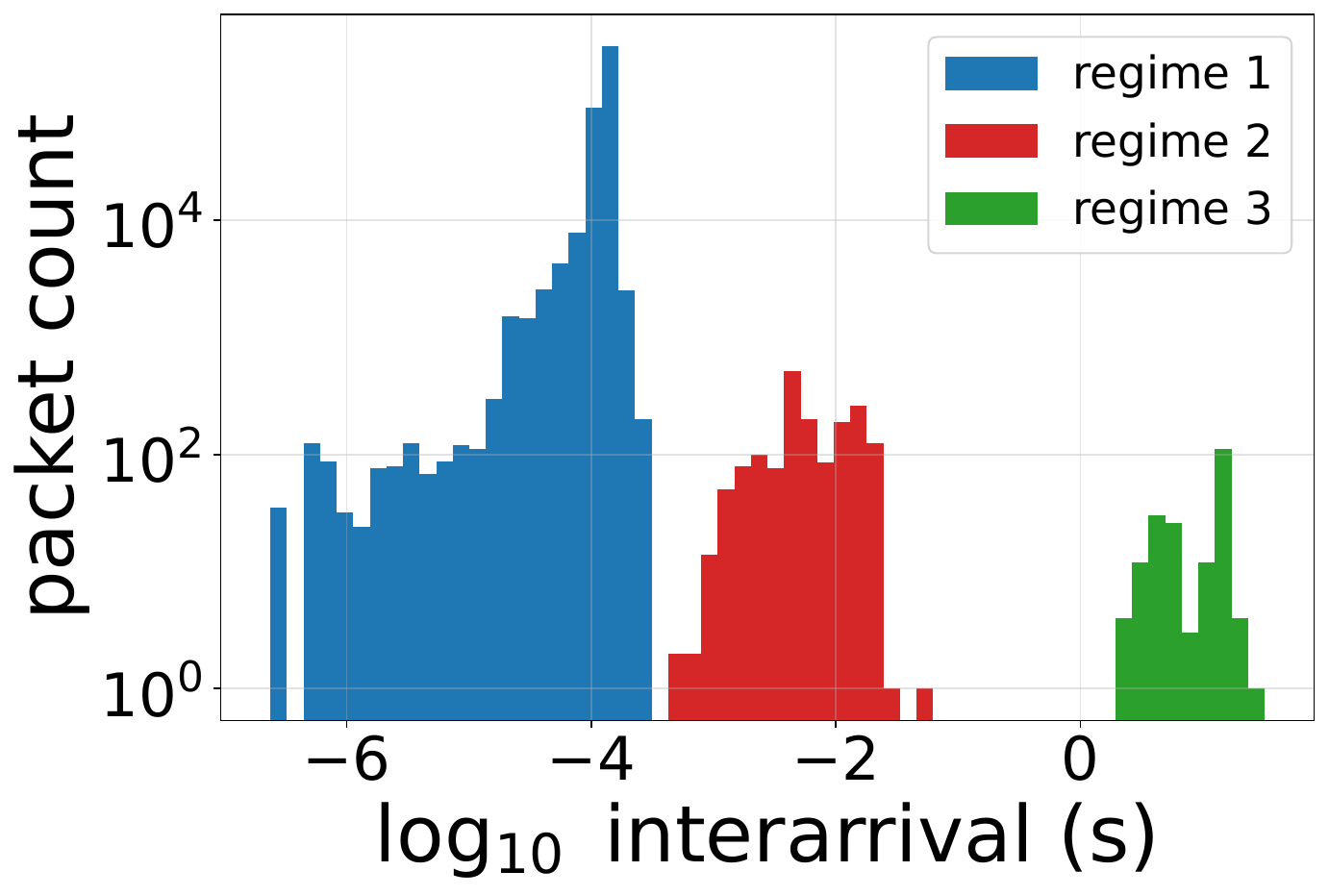}\\

    \end{minipage}

    \vspace{0.7em}

    \includegraphics[width=\linewidth]{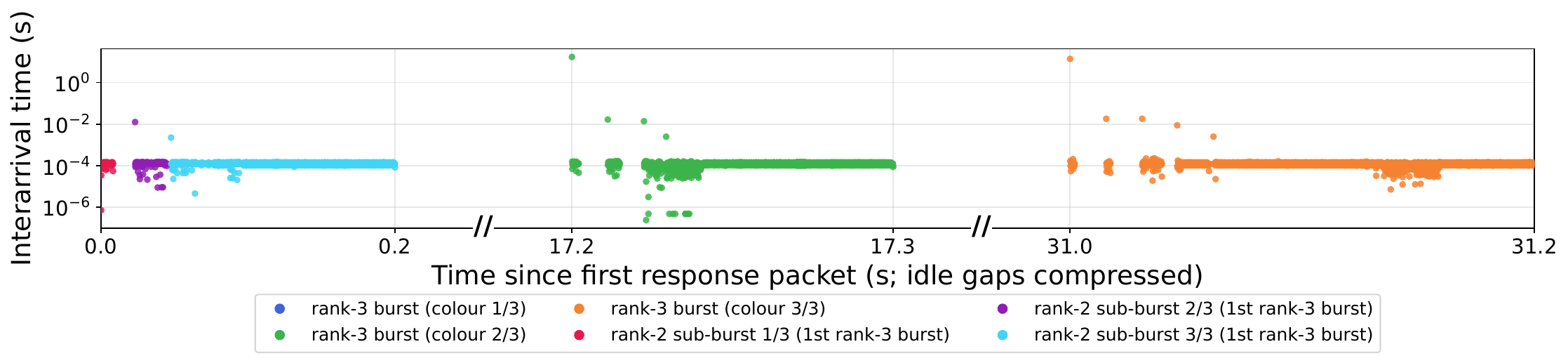}\\

  \caption{
  \textit{Agglomerative hierarchical clustering} (AHC) on text-to-image streaming example. AHC clusters (top left), AHC time-scale burst regimes (top right), example burst classification (bottom).
    }
    \label{fig:streaming_cluster_regimes}
\end{figure}

Network capture shows irregular packet bursts within the same GenAI session, originating from server processing delay and token streaming behavior. We analyzed and validated the GenAI burst structure across network access, and we compared it with traditional file download and video streaming behaviors.

\textbf{Feature Extraction}.
The burst timing structure can be extracted from longer interarrival times that start packet bursts, shown in the IAT-versus-time plot in Figure \ref{fig:streaming_cluster_regimes}. We want to validate application-level packet burst behavior across network access types. Since the Ethernet packet serialization time, $\approx10^{-4}$\,s, is clearly shown in the shortest IAT cluster in the above figure, the other clusters represent application-layer burst timings. Thus, we can use \textit{agglomerative hierarchical clustering (AHC)} \cite{deart2021agglomerative} for data-driven thresholds for clusters. Under Wi-Fi and 5G networks, the separability is less clear, making AHC ineffective and requiring a new method, \textit{throughput burst extraction}.

\textit{Agglomerative Hierarchical Clustering:}  Unlike simple thresholding, which relies on manually selected cutoffs, the AHC method exploits the density and proximity of neighboring log-IAT histogram bins to uncover natural groupings, shown in Figure~\ref{fig:streaming_cluster_regimes}. The number of clusters with the largest silhouette score---a measure of how well each point fits its assigned cluster versus the nearest alternative---sets the number of burst regimes. We identified distinct burst transmission regimes from sub-ms packet serialization to over ten-second server-thinking bursts, ranked by time-scale: R1 near packet serialization, R2 sub-bursts, and R3 bursts. These give data-driven boundaries for packet IAT clusters, where the longer time-scale regimes define the burst IAT (B-IAT) distribution.

\begin{figure}[!tb]
    \centering
    \includegraphics[width=0.8\linewidth]{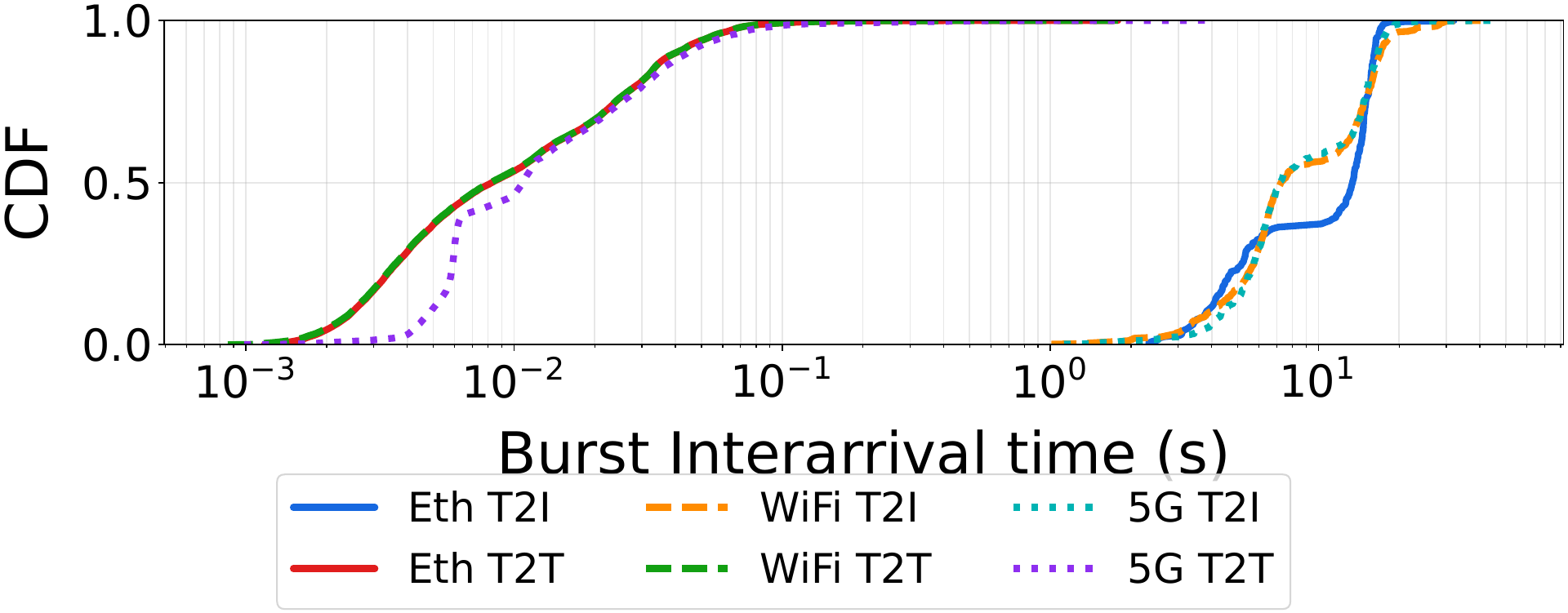}

    \vspace{0.5em}

    \includegraphics[width=0.75\linewidth]{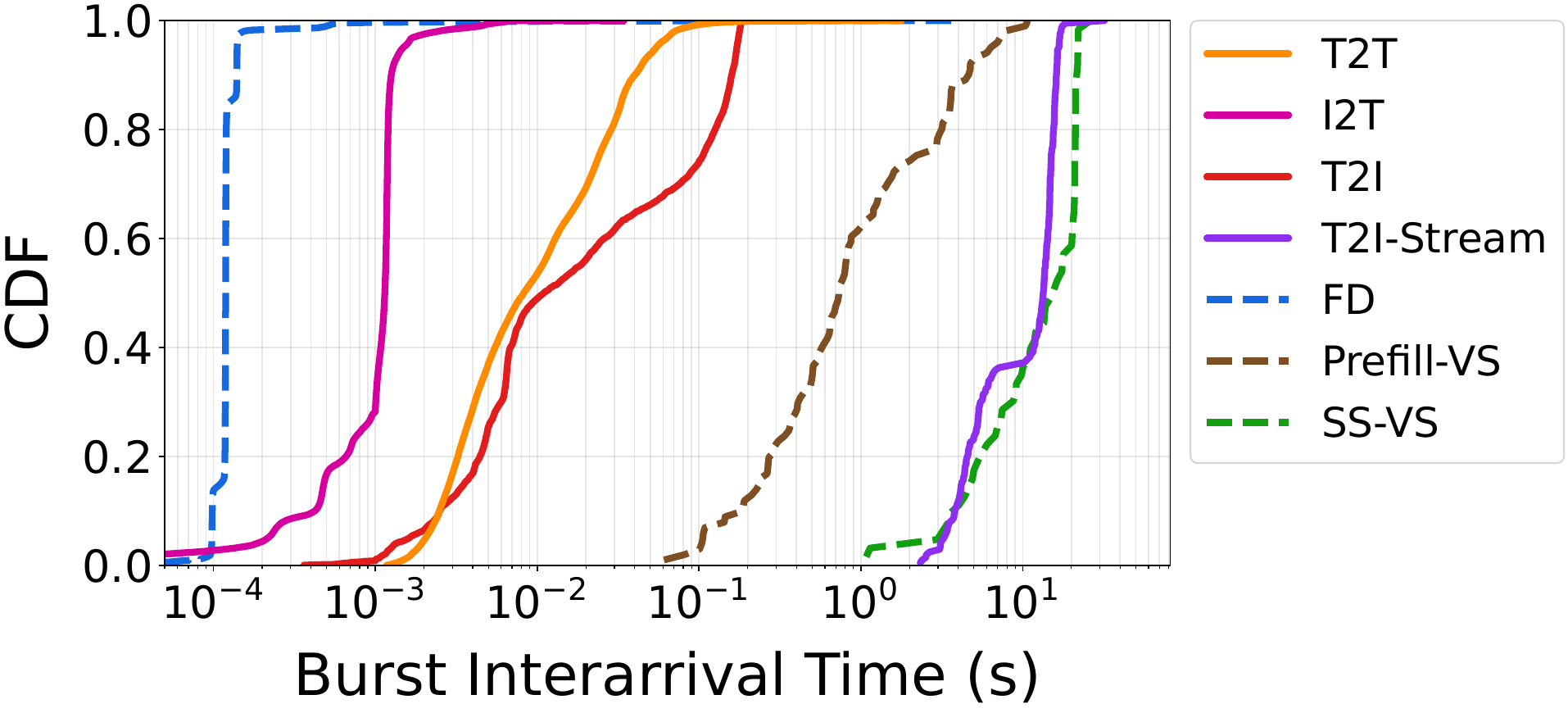}
    \caption{ Burst Interarrival Time (B-IAT) across access networks (top). B-IAT across GenAI and traditional applications (bottom), including file download (FD), and  the prefill and steady-state (SS) video stream (VS).}
    \label{fig:burst_comparison}
\end{figure}

\textit{Throughput Burst Extraction:} Wi-Fi and 5G exhibited longer transmission intervals due to additional wireless access delays, requiring a new \textit{throughput burst extraction} method to extract burst timings. We identify 1\,ms windows with significant throughput and iteratively construct larger bursts from adjacent windows. We uncovered a B-IAT bimodality with a clear separation at 1 second, revealing the largest burst structure above the threshold. We use this method to validate the largest application-layer bursts found by AHC; however, it cannot identify sub-burst structure.

\textit{Burst Behavior Validation across Wi-Fi and 5G:} We validated B-IAT extraction using AHC on Ethernet IAT with the throughput-based burst extraction used on Wi-Fi and 5G data, shown in Figure~\ref{fig:burst_comparison}. Despite Wi-Fi and 5G introducing IAT artifacts in the sub-burst (R2) and serialization (R1) regimes, the throughput-based extraction shows B-IAT distributions aligning closely with AHC on Ethernet data. This validates that the largest-regime IAT distribution is the application B-IAT, independent of network access.

\textbf{Burst Analysis}. GenAI traffic exhibits five types of burstiness, only some of which resemble traditional applications, shown in Figure~\ref{fig:burst_comparison}. First, image upload and non-streaming text (not shown) both send data at a near-constant rate, resembling file download. Second, text streaming has its own B-IAT distribution with a median of 10\,ms, a fraction of the 100\,ms token streaming rate cited in the simple token generator for \cite{li2024eloquent}. Third, partial image streaming in ChatGPT is separated by long server-thinking delays of nondeterministic image generation, which surprisingly arrive at burst intervals similar to deterministic steady-state video buffering. These partial images also contain sub-bursts (R2), shown in Figure~\ref{fig:streaming_cluster_regimes}. Fourth, non-streaming images arrived in bursts spaced by a distribution centered around 10\,ms, rather than the bulk transfer expected for an already generated image. Lastly, code generation demonstrated diverse burstiness in both upload and download within one capture: periods of file-download bursts, structured back-and-forth exchanges, and highly variable burstiness at the end. We keep code generation as a qualitative comparison and leave modeling such complex usage to future work.

\subsection{ Modeling and ns-3 Simulation}

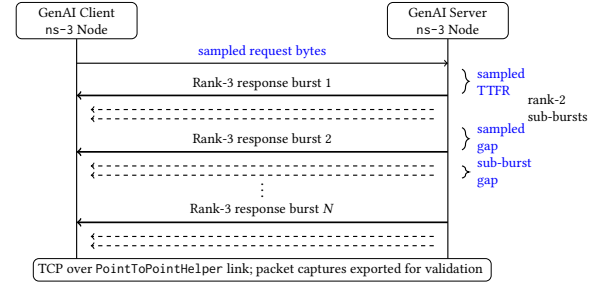
\begin{figure}[t]
    \centering
    \begin{adjustbox}{max width=0.9\linewidth}
    \begin{tikzpicture}[
        node distance=1cm,
        box/.style={draw, rounded corners, align=center, minimum width=3.0cm, minimum height=0.9cm},
        arrow/.style={->, thick},
        burst/.style={->, very thick},
        subburst/.style={->, dashed, thick},
        note/.style={align=left, font=\normalsize},
        every node/.style={font=\normalsize}
    ]

    \node[box] (client) {GenAI Client\\\texttt{ns-3} Node};
    \node[box, right=6.2cm of client] (server) {GenAI Server\\\texttt{ns-3} Node};

    \draw[thick] (client.south) -- ++(0,-5.45);
    \draw[thick] (server.south) -- ++(0,-5.45);

    \coordinate (c0) at ($(client.south)+(0,-0.6)$);
    \coordinate (s0) at ($(server.south)+(0,-0.6)$);

    \draw[arrow] (c0) -- node[above]{\textcolor{blue!100!black}{sampled request bytes}} (s0);

    \draw[decorate, decoration={brace, amplitude=5pt}, thick]
        ($(server.south)+(0.35,-0.75)$) -- ($(server.south)+(0.35,-1.35)$)
    node[midway, right=7pt, align=left]{\textcolor{blue!100!black}{sampled}\\\textcolor{blue!100!black}{TTFR}};
    \coordinate (s1) at ($(server.south)+(0,-1.4)$);
    \coordinate (c1) at ($(client.south)+(0,-1.4)$);
    \draw[burst] (s1) -- node[above]{Rank-3 response burst 1} (c1);

    \draw[subburst] ($(server.south)+(-0.35,-1.75)$) -- ($(client.south)+(0.35,-1.75)$);
    \draw[subburst] ($(server.south)+(-0.35,-1.98)$) -- ($(client.south)+(0.35,-1.98)$);
    \node[note, right=0.35cm of server, yshift=-2.15cm] {rank-2\\sub-bursts};

    \draw[decorate, decoration={brace, amplitude=5pt}, thick]
        ($(server.south)+(0.35,-2.2)$) -- ($(server.south)+(0.35,-2.75)$)
    node[midway, right=7pt, align=left]{\textcolor{blue!100!black}{sampled}\\\textcolor{blue!100!black}{gap}};

    \coordinate (s2) at ($(server.south)+(0,-2.8)$);
    \coordinate (c2) at ($(client.south)+(0,-2.8)$);
    \draw[burst] (s2) -- node[above]{Rank-3 response burst 2} (c2);

    \draw[decorate, decoration={brace, amplitude=5pt}, thick]
    ($(server.south)+(0.35,-3.15)$) -- ($(server.south)+(0.35,-3.45)$)
    node[midway, right=7pt, align=left]{\textcolor{blue!100!black}{sub-burst}\\\textcolor{blue!100!black}{gap}};

    \draw[subburst] ($(server.south)+(-0.35,-3.15)$) -- ($(client.south)+(0.35,-3.15)$);
    \draw[subburst] ($(server.south)+(-0.35,-3.38)$) -- ($(client.south)+(0.35,-3.38)$);

    \node at ($(client.south)!0.5!(server.south)+(0,-3.65)$) {\Large $\vdots$};

    \coordinate (sN) at ($(server.south)+(0,-4.55)$);
    \coordinate (cN) at ($(client.south)+(0,-4.55)$);
    \draw[burst] (sN) -- node[above]{Rank-3 response burst $N$} (cN);

    \draw[subburst] ($(server.south)+(-0.35,-4.90)$) -- ($(client.south)+(0.35,-4.90)$);
    \draw[subburst] ($(server.south)+(-0.35,-5.13)$) -- ($(client.south)+(0.35,-5.13)$);

    \node[
        draw,
        rounded corners,
        align=center,
        font=\normalsize,
        below=5.9cm of $(client)!0.5!(server)$,
        minimum width=7.0cm
    ] {
        TCP over \texttt{PointToPointHelper} link; packet captures exported for validation
    };

    \end{tikzpicture}
    \end{adjustbox}

    \caption{ ns-3 simulation burst structure, setting distributions for time-to-first-response (TTFR), and burst parameters per modality. 
    }
    \label{fig:ns3_ranked_burst_timing}
\end{figure}

\begin{table*}[!tb]
    \centering
    \caption{Selected distribution models for different GenAI traffic burst structure, compared by Akaike/Bayesian Information Criterion (AIC/BIC) across candidate distributions. Best-fit, least-complex model minimizing AIC/BIC for different $R$ burst regimes. (NBN: negative binomial; Pois.: Poisson; LN: lognormal; Par.: Pareto; $K$-logGMM: $K$-component log Gaussian mixture model.)}
    \label{tab:genai_distribution_models}
    \renewcommand{\arraystretch}{1.15}
    \small
    \begin{tabular}{lcccccccc}
        \hline
        \textbf{Modality} &
        \textbf{Prompt} &
        \textbf{Response} &
        \textbf{TTFR} &
        \textbf{\# R3} &
        \textbf{R3 IAT} &
        \textbf{\# R2} &
        \textbf{R2 IAT} &
        \textbf{R2 Bytes} \\
        \hline

        Text-to-Text Streaming      
        & LN   
        & Burst   
        & 2-logGMM
        & -- 
        & --        
        & NBN   
        & 3-logGMM 
        & 2-logGMM \\

        Text-to-Text Non-Streaming  
        & LN   
        & Par. 
        & 2-logGMM
        & -- 
        & --        
        & --   
        & -- 
        & -- \\

        Image-to-Text               
        & LN   
        & LN   
        & 2-logGMM
        & -- 
        & --        
        & --    
        & --       
        & -- \\

        Text-to-Image Streaming     
        & Par. 
        & Burst   
        & 2-logGMM
        & Constant 
        & 3-logGMM  
        & Pois. 
        & 3-logGMM 
        & 1-logGMM \\

        Text-to-Image Non-Streaming 
        & Par. 
        & Burst   
        & 2-logGMM
        & -- 
        & --        
        & NBN   
        & 3-logGMM 
        & LN \\

        \hline
    \end{tabular}
\end{table*}

Although some GenAI burst behavior resembles traditional applications, no comprehensive GenAI model exists. We build the first mechanistic model of single-session GenAI traffic---fitting distributions to prompt size, TTFR, and the response burst structure---and implement it as a client/server application pair in ns-3. Simulated traffic reproduces empirical burst interarrival times within 2--25\% normalized Wasserstein distance, whereas a constant token-streaming baseline deviates by up to 491\%. We describe the model, its ns-3 implementation, and its validation in turn.

\textbf{Modeling:} We probabilistically model each burst timing regime identified by AHC on Ethernet data. A burst of rank $R$ comprises $N_R$ sub-bursts, each of byte size $B_R$, spaced by B-IAT $\Delta t_{B,R}$ seconds. For bursts (rank-3) and sub-bursts (rank-2), we fit each variable against candidate distributions common in networking, such as log-normal and Pareto (Table~\ref{tab:genai_distribution_models}), selecting the distribution that maximizes goodness-of-fit by Akaike Information Criterion (AIC) while penalizing model complexity by Bayesian Information Criterion (BIC) to prevent overfitting. TTFR fits a 2- or 3-component log-Gaussian mixture model in all cases.

\textbf{ns-3 Implementation}. We implement the burst model in the widely used open-source ns-3 simulator~\cite{maza2016framework} as a client/server application pair, \texttt{GenAIUser} and \texttt{GenAIServer}, shown in Figure~\ref{fig:ns3_ranked_burst_timing}. The applications open a single TCP connection, over which the client sends a prompt request. The server reassembles the request, samples a processing delay for the TTFR packet, and then generates $N_3$ rank-3 bursts; each burst may comprise rank-2 sub-bursts, sampling $N_2$ sub-bursts of $B_2$ bytes spaced by $\Delta t_{B,2}$. A \texttt{JSON} configuration file exposes knobs for every random-variable distribution, and each simulation is deterministic and reproducible from its ns-3 RNG seed. The model represents burst reception as seen by a single user; the application pair enables future research such as multi-user network sessions. As a baseline, we also implement the token-streaming model of Eloquent~\cite{li2024eloquent}, which sends one sub-MTU token every 100\,ms.

\textbf{Simulation Setup}. We run both MINT's GenAI model and the baseline on two ns-3 nodes, client and server, joined by a point-to-point link mimicking the empirical test setup: 1500-byte MTU, 100\,Mbps data rate, and 5\,ms round-trip time. We capture 200 client-side simulated PCAP traces.

\textbf{Evaluation Method}. We evaluate whether MINT's simulated traffic preserves the empirical response-size and B-IAT distributions, using normalized Wasserstein distance (WS), a metric used to compare network simulators~\cite{ yang2022deepqueuenet}. WS measures the average distribution shift needed to match cumulative distribution functions, expressed as a percentage of the mean; lower is better. We divide the empirical PCAPs in half into training and validation sets, then compare the training set against 1) the empirical validation set, 2) samples from the fitted distributions, 3) MINT ns-3 simulated PCAPs, and 4) baseline token-generator PCAPs.

\begin{figure}[t]
    \centering
    \includegraphics[width=0.75\linewidth]{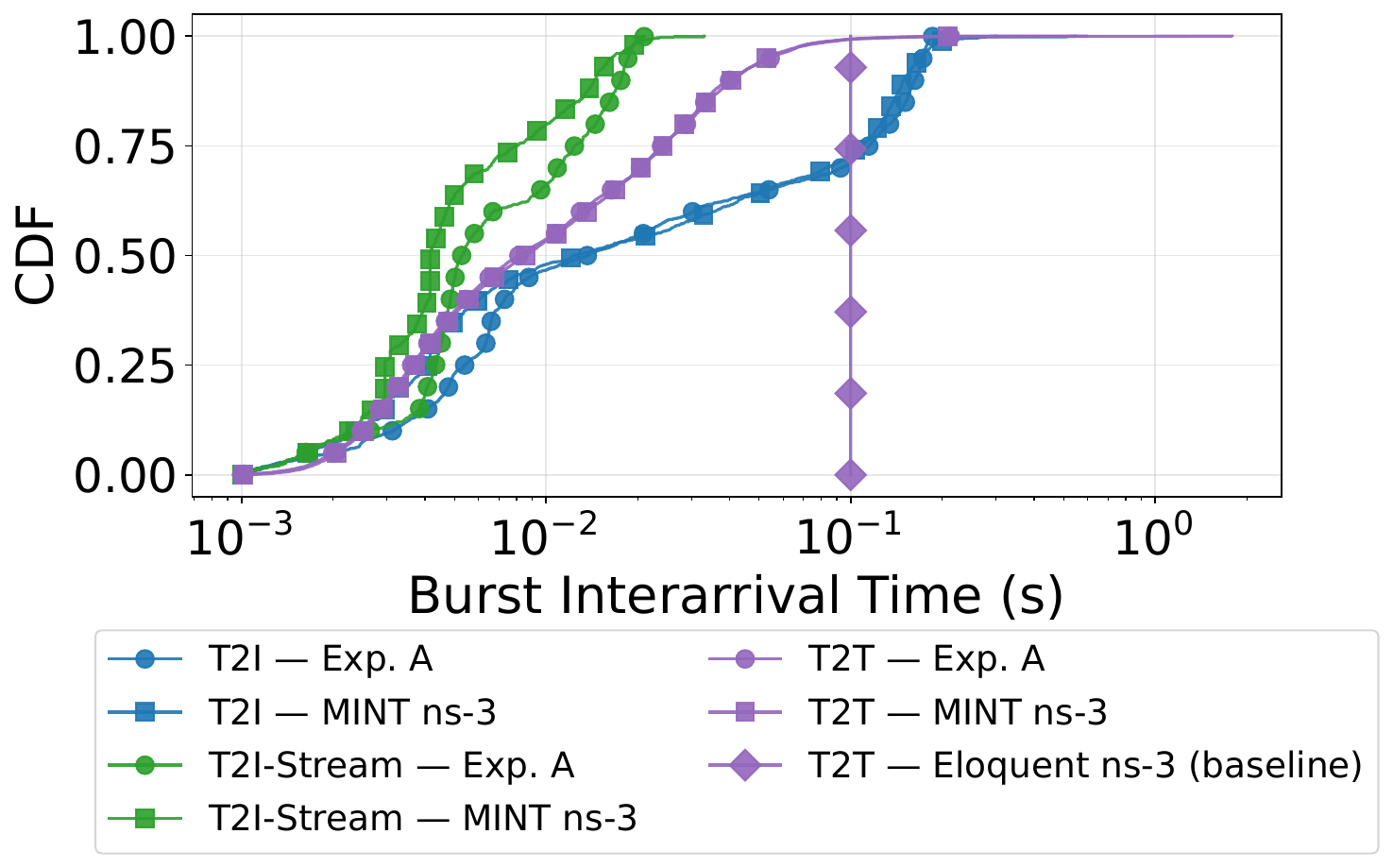}\\

    \caption{Burst interarrival times comparing training with MINT ns-3 and baseline token streaming in ns-3 (Eloquent \cite{li2024eloquent}).}
    \label{fig:validation_response_distributions}
\end{figure}

\textbf{Burst Distribution Validation}. MINT's simulated traffic matches the empirical burst behavior closely---within 2.3--9\% WS for B-IAT across modalities and under 5\% for total response size, versus up to 491\% for the baseline---with one exception we examine below (Table~\ref{tab:validation_ks_ws}). All comparisons are made against the empirical validation set, itself within 6\% WS of training data. First, the constant token generator~\cite{li2024eloquent}, designed to validate token transmission rather than packet-level bursts, cannot reflect the variation in empirical T2T burst timing: its B-IAT WS reaches 491\%, whereas MINT's simulation and fitted distributions stay within 4.6\%. Second, the exception: the text-to-image streaming model for rank-2 sub-bursts reaches 24.3\% WS; Figure~\ref{fig:validation_response_distributions} shows we capture the bimodal structure of sub-bursts but not their full distribution. Overall, the ns-3 simulator reproduces empirical burst behavior across modalities and improves on the constant token-generation model by two orders of magnitude.

\begin{table}[t]
    \centering
    \caption{Validation between empirical training set (Emp.~A) and the validation set (Emp.~B), fitted distributions (Dist.), MINT ns-3 simulation, and the Eloquent baseline~\cite{li2024eloquent} using  normalized Wasserstein distance relative to the mean (WS).}
    \label{tab:validation_ks_ws}
    \renewcommand{\arraystretch}{1.15}
    \small
    \setlength{\tabcolsep}{3.5pt}
    \begin{tabular}{llcccc}
        \toprule
        \textbf{Modality} &
        \textbf{Metric} &
        \multicolumn{4}{c}{\textbf{WS (\%)}} \\
        \cmidrule(lr){3-6}
        \multicolumn{2}{c}{\textbf{Emp. A vs.}} &
        \textbf{Emp. B} &
        \textbf{Dist.} &
        \textbf{MINT} &
        \textbf{Eloquent} \\
        \midrule
        T2T-S & Burst IAT        & 2.36 & 4.55 & 4.27 & 491 \\
        \midrule
        T2I-S & Rank-2 Burst IAT & 4.23 & 5.56 & 24.3 & -- \\
        \midrule
        T2I   & Burst IAT        & 5.41 & 7.52  & 9.04   & -- \\
        \bottomrule
    \end{tabular}
\end{table}

\section{Conclusion}This paper presented MINT, the first measurement-driven network-simulator model of GenAI traffic, which we open-source\footref{fn:ns3} together with its dataset\footref{fn:data}. Across three providers and four modalities, we showed that GenAI traffic is not simply download-heavy, but modality-dependent and characterized by multi-timescale burst structure distinct from bulk transfer and video streaming. We reproduced this behavior with an ns-3 application pair driven by fitted empirical distributions, capturing the timing variability in packets that constant-rate baseline in \cite{li2024eloquent} does not model. Across modalities, simulated burst distributions matched empirical measurements within 2--25\% normalized Wasserstein distance.

\textbf{Limitations and Future Work}. MINT currently models single-shot text and image API interactions. It does not capture browser/app traffic, sequential prompting with human think time, or automated agentic workloads. Because our measurements are packet-level and encrypted, application-stage boundaries are inferred from traffic structure rather than directly observed from decrypted traffic, such as MITM. We will later extend MINT to browser/app, agentic  multi-turn workloads, including code-generation and agentic traffic, and use decrypted measurements, such as controlled MITM interception, to validate application-stage boundaries. We also plan to evaluate network optimization techniques under realistic simulated GenAI workloads.


\bibliographystyle{ACM-Reference-Format}
\bibliography{acmicm2026/acmpaper/refs/andrew,acmicm2026/acmpaper/refs/samson}

@inproceedings{feng2026beamformer,
  title={BeamFormer: Transformer-based Beam Management for 6G Networks},
  author={Feng, Shunqiang and Kanjilal, Swastik and Qian, Kun and Jain, Ish Kumar},
  booktitle={Proceedings of the 24th Annual International Conference on Mobile Systems, Applications and Services},
  pages={652--666},
  year={2026}
}

@article{jin2025end,
  title={End-to-End Coordination of {RAN} and Edge Server for Latency-Critical Inference Serving over Cellular Networks},
  author={Jin, Sunghyun and Kim, Serae and Ha, Sangtae and Lee, Kyunghan},
  journal={Proceedings of the ACM on Networking},
  volume={3},
  number={CoNEXT4},
  pages={1--23},
  year={2025},
  publisher={ACM New York, NY, USA}
}

@inproceedings{li2024eloquent,
  title={Eloquent: A More Robust Transmission Scheme for {LLM} Token Streaming},
  author={Li, Hanchen and Liu, Yuhan and Cheng, Yihua and Ray, Siddhant and Du, Kuntai and Jiang, Junchen},
  booktitle={Proceedings of the 2024 SIGCOMM Workshop on Networks for {AI} Computing},
  pages={34--40},
  year={2024}
}

@inproceedings{wang2023diffusiondb,
  title={{DiffusionDB}: A Large-scale Prompt Gallery Dataset for Text-to-Image Generative Models},
  author={Wang, Zijie J and Montoya, Evan and Munechika, David and Yang, Haoyang and Hoover, Benjamin and Chau, Duen Horng},
  booktitle={Proceedings of the 61st annual meeting of the association for computational linguistics (volume 1: Long papers)},
  pages={893--911},
  year={2023}
}

@article{wang2024revisiting,
  title={Revisiting Service Level Objectives and System Level Metrics in Large Language Model Serving},
  author={Wang, Zhibin and Li, Shipeng and Zhou, Yuhang and Li, Xue and Zhang, Zhonghui and Cam-Tu, Nguyen and Gu, Rong and Tian, Chen and Chen, Guihai and Zhong, Sheng},
  journal={arXiv preprint arXiv:2410.14257},
  year={2024}
}

@article{paxson2002wide,
  title={{Wide-Area} Traffic: The Failure of Poisson Modeling},
  author={Paxson, Vern and Floyd, Sally},
  journal={IEEE/ACM Transactions on networking},
  volume={3},
  number={3},
  pages={226--244},
  year={2002},
  publisher={IEEE}
}

@article{willinger2019lessons,
  title={Lessons from "On the Self-Similar Nature of {Ethernet} Traffic"},
  author={Willinger, Walter and Taqqu, Murad S and Wilson, Daniel V},
  journal={ACM SIGCOMM Computer Communication Review},
  volume={49},
  number={5},
  pages={56--62},
  year={2019},
  publisher={ACM New York, NY, USA}
}

@inproceedings{bojovic2022enabling,
  title={Enabling {NGMN} Mixed Traffic Models for {ns-3}},
  author={Bojovic, Biljana and Lagen, Sandra},
  booktitle={Proceedings of the 2022 Workshop on {ns-3}},
  pages={127--134},
  year={2022}
}

@techreport{wong2006comments,
  author      = {Wendy C. Wong and Roshni Srinivasan and Hannah Hyunjeong Lee and Kerstin Johnsson and Jerry Sydir and Sassan Ahmadi and Belal Hamzeh and Shailender Timiri and I-Kang Fu and Peter Wang and David Chen and Hua Xu and Roger Peterson and Aik Chindapol and Teck Hu and Mike Hart and Sunil Vadgama and Peng-Yong Kong and Haiguang Wang and Dharma Basgeet and Yong Sun and Jun Bae Ahn and Hyunjeong Kang and Jaeweon Cho and Hyoungkyu Lim},
  title       = {{Comments and Proposal to Replace Traffic Models in IEEE 802.16j-06/013}},
  institution = {{IEEE 802.16 Broadband Wireless Access Working Group}},
  number      = {{IEEE C802.16j-06/093r3}},
  year        = {2006},
  month       = sep,
  day         = {26},
  type        = {IEEE 802.16 Contribution}
}

@misc{ericsson2025genai,
  author       = {{Ericsson}},
  title        = {{GenAI Data Traffic Today}},
  howpublished = {\url{https://www.ericsson.com/en/reports-and-papers/mobility-report/articles/genai-data-traffic-today-june-2025}},
  year         = {2025},
  month        = jun,
  note         = {Ericsson Mobility Report, accessed April 24, 2026}
}

@misc{cisco2018vni,
  author       = {{Cisco}},
  title        = {{Cisco Visual Networking Index: Forecast and Trends, 2017--2022}},
  howpublished = {\url{https://www.cisco.com/c/dam/m/en_us/solutions/service-provider/vni-forecast-highlights/pdf/Global_Device_Growth_Traffic_Profiles.pdf}},
  year         = {2018},
  note         = {Accessed April 24, 2026}
}

@article{loh2022youtube,
  title={{YouTube} Dataset on Mobile Streaming for Internet Traffic Modeling and Streaming Analysis},
  author={Loh, Frank and Wamser, Florian and Poign{\'e}e, Fabian and Gei{\ss}ler, Stefan and Ho{\ss}feld, Tobias},
  journal={Scientific Data},
  volume={9},
  number={1},
  pages={293},
  year={2022},
  publisher={Nature Publishing Group UK London}
}

@inproceedings{tagami2026understanding,
  title        = {Understanding Network Impact of Generative AI: Application-Level Traffic Measurement},
  author       = {Tagami, Atsushi and Sekigawa, Shu and Ueda, Kazuaki and Fukumoto, Norihiro and Sasaki, Chikara},
  booktitle    = {Proceedings of IEEE INFOCOM 2026 Workshop on 6G AI-RAN},
  year         = {2026},
  organization = {IEEE}
}

@inproceedings{maza2016framework, 
title={A Framework for Generating {HTTP} Adaptive Streaming Traffic in {ns-3}}, 
author={Maza, William David Diego}, 
booktitle={SIMUTools-9th E{AI} International Conference on Simulation Tools and Techniques-2016}, 
year={2016} 
}

@article{alhazbi2025llms,
  title={{LLMs} Have Rhythm: Fingerprinting Large Language Models Using Inter-Token Times and Network Traffic Analysis},
  author={Alhazbi, Saeif and Hussain, Ahmed and Oligeri, Gabriele and Papadimitratos, Panos},
  journal={IEEE Open Journal of the Communications Society},
  year={2025},
  publisher={IEEE}
}

@inproceedings{deart2021agglomerative,
  title={Agglomerative Clustering of Network Traffic Based on Various Approaches to Determining the Distance Matrix},
  author={Deart, Vladimir and Mankov, Vladimir and Krasnova, Irina},
  booktitle={2021 28th Conference of Open Innovations Association (FRUCT)},
  pages={81--88},
  year={2021},
  organization={IEEE}
}

@inproceedings{xiao2025streaming,
  title={Streaming, Fast and Slow: Cognitive Load-Aware Streaming for Efficient {LLM} Serving},
  author={Xiao, Chang and Yang, Zixiaofan},
  booktitle={Proceedings of the 38th Annual ACM Symposium on User Interface Software and Technology},
  pages={1--13},
  year={2025}
}

@article{koneva2025introducing,
  title={Introducing Large Language Models as the Next Challenging Internet Traffic Source},
  author={Koneva, Nataliia and Navarro, Alejandro Leonardo Garc{\'\i}a and S{\'a}nchez-Maci{\'a}n, Alfonso and Hern{\'a}ndez, Jos{\'e} Alberto and Zukerman, Moshe and de Dios, {\'O}scar Gonz{\'a}lez},
  journal={arXiv preprint arXiv:2504.10688},
  year={2025}
}

@article{montieri2026prompts,
  title={From Prompts to Packets: A View from the Network on {ChatGPT}, {Copilot}, and {Gemini}},
  author={Montieri, Antonio and Nascita, Alfredo and Pescap, Antonio},
  journal={Computer Networks},
  pages={112237},
  year={2026},
  publisher={Elsevier}
}

@inproceedings{cheng2025hello,
  title={Hello, {GenAI}? Dissecting Human to Generative {AI} Calling},
  author={Cheng, Ruizhi and Pathak, Surendra and Xie, Guowu and Varvello, Matteo and Chen, Songqing and Han, Bo},
  booktitle={Proceedings of the 2025 ACM Internet Measurement Conference},
  pages={308--324},
  year={2025}
}

@inproceedings{yang2022deepqueuenet,
  title={{DeepQueueNet}: Towards Scalable and Generalized Network Performance Estimation with Packet-Level Visibility},
  author={Yang, Qingqing and Peng, Xi and Chen, Li and Liu, Libin and Zhang, Jingze and Xu, Hong and Li, Baochun and Zhang, Gong},
  booktitle={Proceedings of the ACM SIGCOMM 2022 Conference},
  pages={441--457},
  year={2022}
}

@inproceedings{qu2025tokenflow,
  title={{TokenFlow}: Unified Image Tokenizer for Multimodal Understanding and Generation},
  author={Qu, Liao and Zhang, Huichao and Liu, Yiheng and Wang, Xu and Jiang, Yi and Gao, Yiming and Ye, Hu and Du, Daniel K and Yuan, Zehuan and Wu, Xinglong},
  booktitle={Proceedings of the Computer Vision and Pattern Recognition Conference},
  pages={2545--2555},
  year={2025}
}

@article{sivaroopan2025comprehensive,
  title={A Comprehensive Survey on Network Traffic Synthesis: From Statistical Models to Deep Learning},
  author={Sivaroopan, Nirhoshan and Silva, Kaushitha and Madarasingha, Chamara and Dahanayaka, Thilini and Jourjon, Guillaume and Jayasumana, Anura and Thilakarathna, Kanchana},
  journal={arXiv preprint arXiv:2507.01976},
  year={2025}
}

@article{zink2009characteristics,
  title={Characteristics of {YouTube} Network Traffic at a Campus Network--Measurements, Models, and Implications},
  author={Zink, Michael and Suh, Kyoungwon and Gu, Yu and Kurose, Jim},
  journal={Computer networks},
  volume={53},
  number={4},
  pages={501--514},
  year={2009},
  publisher={Elsevier}
}

@misc{openai2025enterpriseai,
  author       = {{OpenAI}},
  title        = {The State of Enterprise AI 2025},
  year         = {2025},
  howpublished = {\url{https://openai.com/business/guides-and-resources/the-state-of-enterprise-ai-2025-report/}},
  note         = {Accessed: 2026-08-29}
}

@inproceedings{patil2025bfcl,
  title={The Berkeley Function Calling Leaderboard (BFCL): From Tool Use to Agentic Evaluation of Large Language Models},
  author={Patil, Shishir G. and Mao, Huanzhi and Cheng-Jie Ji, Charlie and Yan, Fanjia and Suresh, Vishnu and Stoica, Ion and E. Gonzalez, Joseph},
  year={2024},
  booktitle = {Advances in Neural Information Processing Systems},
}

@techreport{stanford2026AI,
  author      = {Landay, James and Lyons, Terah and Manyika, James and Niebles, Juan Carlos and Shoham, Yoav and Tabassi, Elham and Wald, Russell and Walsh, Toby and Weld, Dan},
  title       = {The AI Index 2026 Annual Report},
  institution = {AI Index Steering Committee, Institute for Human-Centered AI, Stanford University},
  year        = {2026},
  month       = {April},
  address     = {Stanford, CA},
  url         = {https://hai.stanford.edu/ai-index/2026-ai-index-report}
}

@article{yang2025qwen3,
  title={Qwen3 technical report},
  author={Yang, An and Li, Anfeng and Yang, Baosong and Zhang, Beichen and Hui, Binyuan and Zheng, Bo and Yu, Bowen and Gao, Chang and Huang, Chengen and Lv, Chenxu and others},
  journal={arXiv preprint arXiv:2505.09388},
  year={2025}
}

@misc{nyc-taxi-trip-duration,
    author = {Meg Risdal},
    title = {New York City Taxi Trip Duration},
    year = {2017},
    howpublished = {\url{https://kaggle.com/competitions/nyc-taxi-trip-duration}},
    note = {Kaggle}
}

@inproceedings{zhang2024quic,
  title={Quic is not quick enough over fast internet},
  author={Zhang, Xumiao and Jin, Shuowei and He, Yi and Hassan, Ahmad and Mao, Z Morley and Qian, Feng and Zhang, Zhi-Li},
  booktitle={Proceedings of the ACM Web Conference 2024},
  pages={2713--2722},
  year={2024}
}
\newpage 

\clearpage
\section{Appendix}

\begin{figure}[H]
    \centering
    \includegraphics[width=\linewidth]{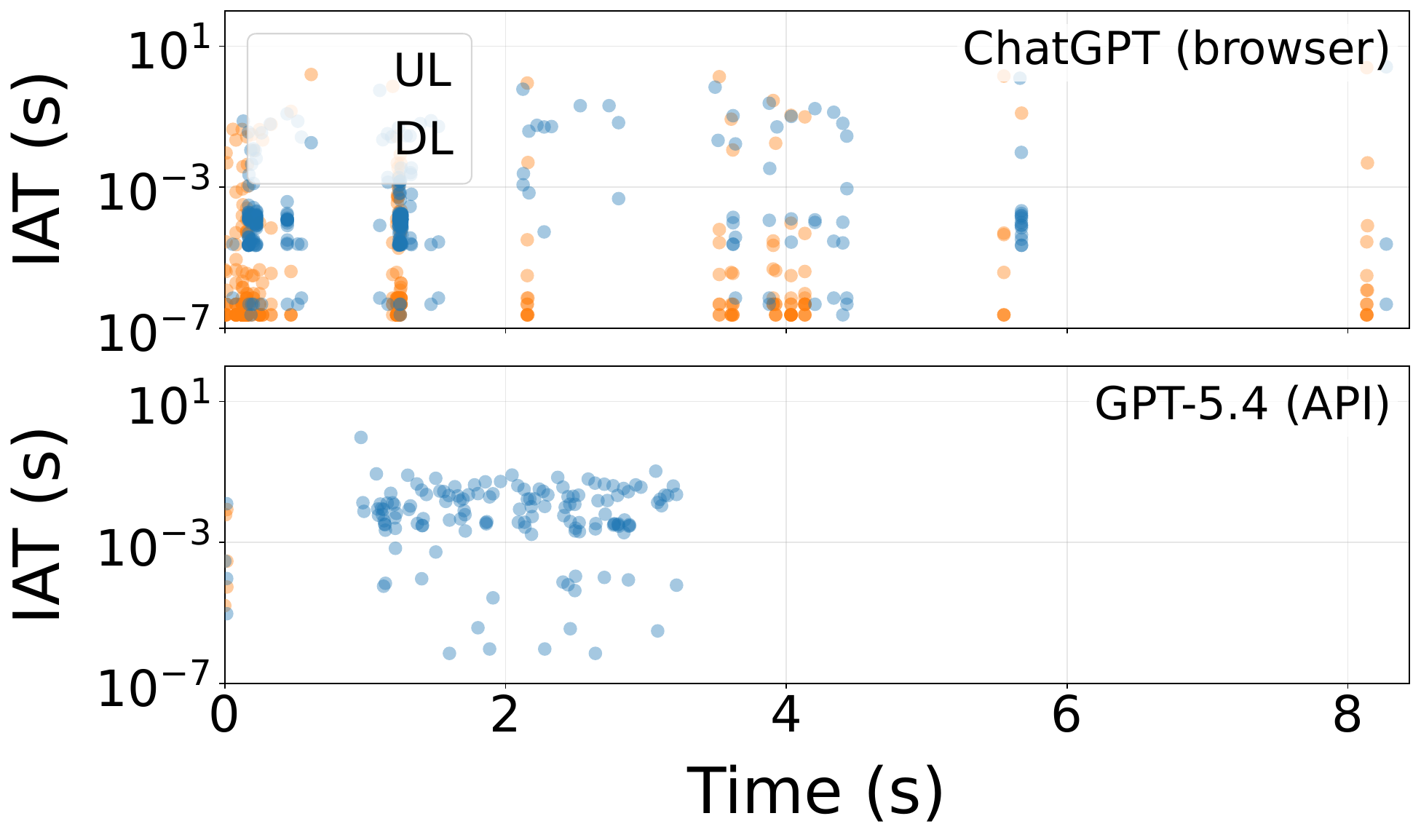}
    \caption{Browser vs API Interarrival Time (IAT) Bursts}
    \label{fig:browser_vs_api}
\end{figure}

\begin{figure}[H]
    \centering
    \includegraphics[width=\linewidth]{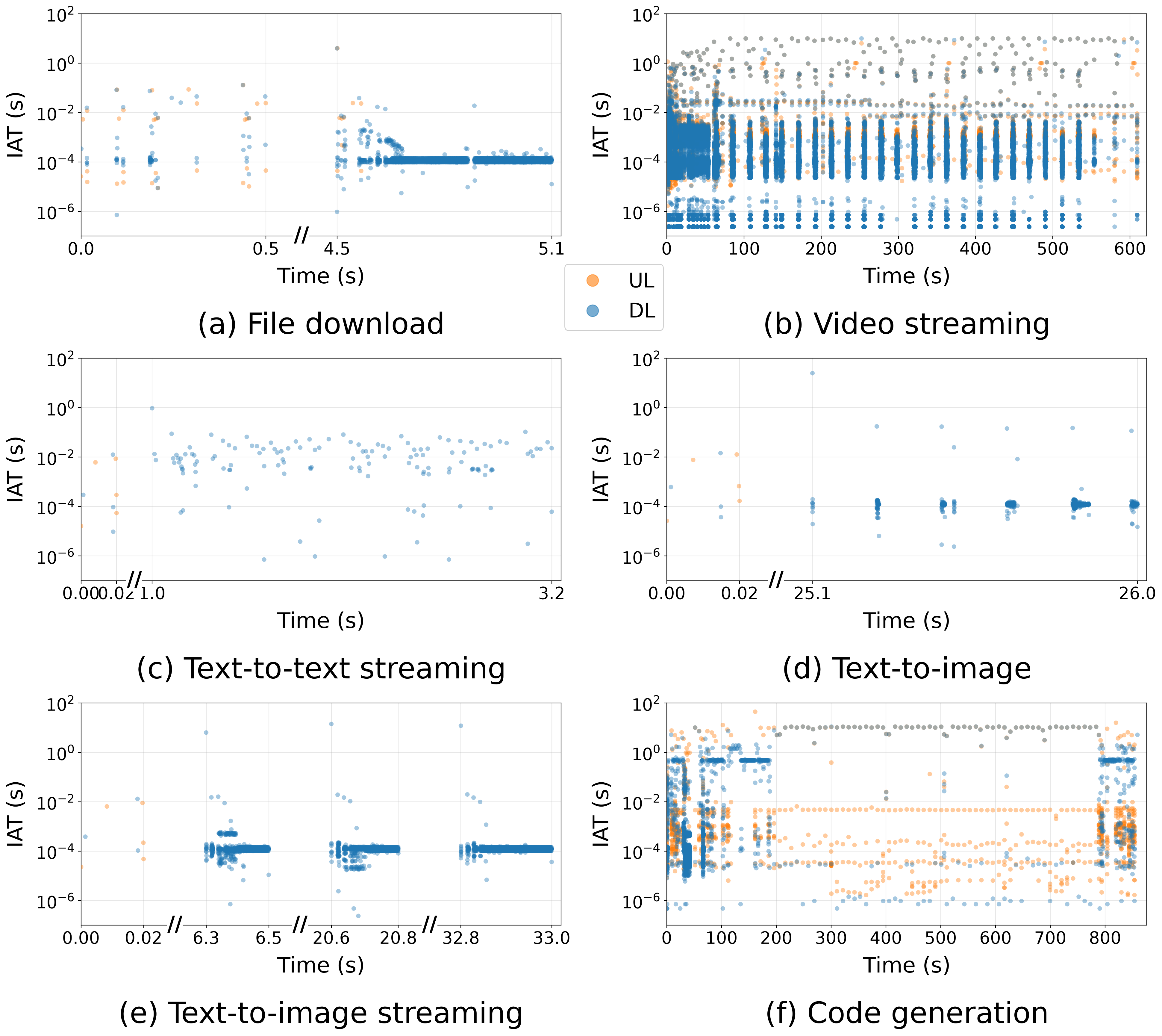}
    \caption{Interarrival Time Comparison between traditional and GenAI applications.}
    \label{fig:burst_comparison}
\end{figure}

\begin{figure}[H]
    \centering
    \includegraphics[width=\linewidth]{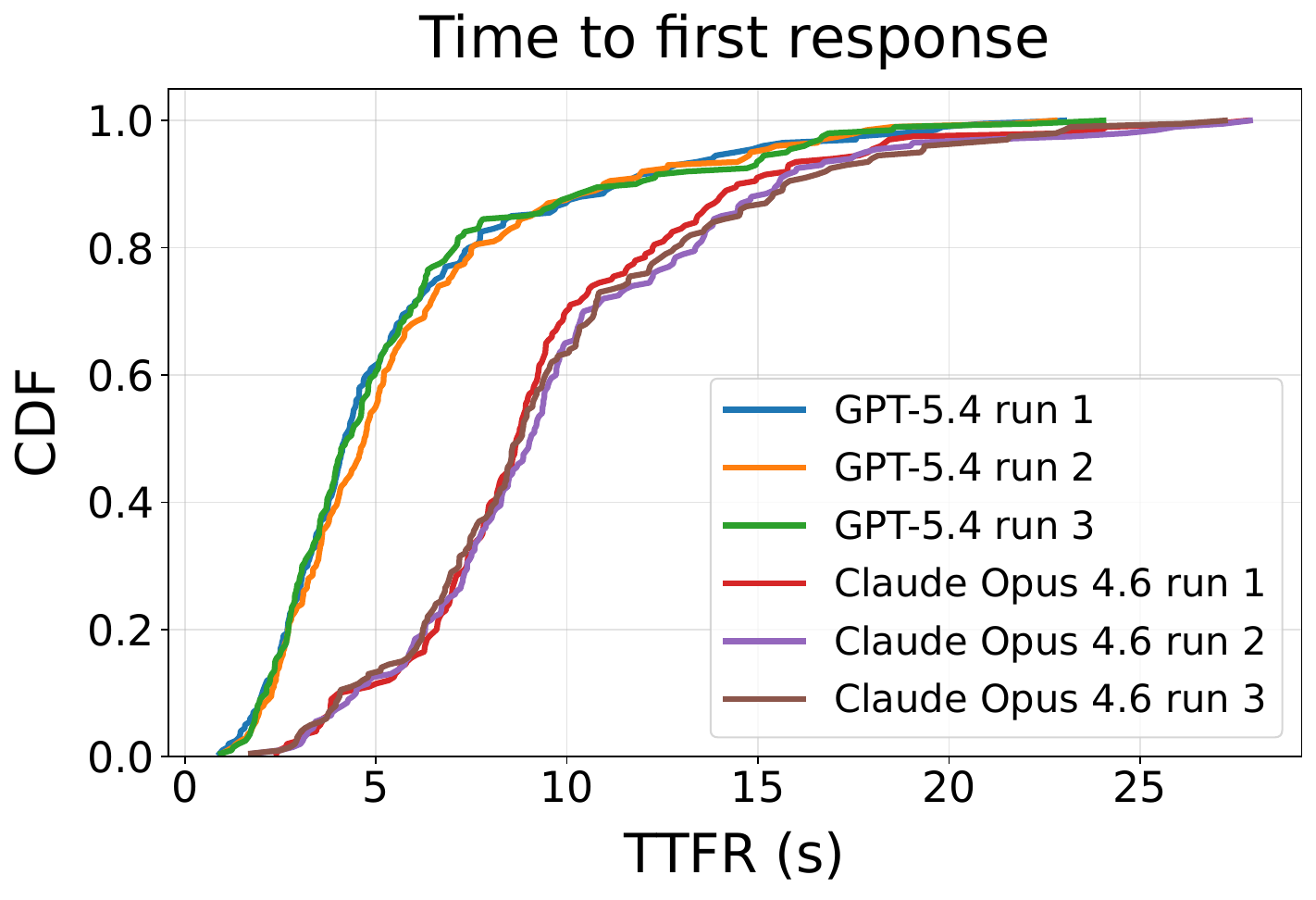}
    \caption{Distribution of time to first response across repeated runs.}
    \label{fig:repeatability}
\end{figure}

\begin{figure}[H]
    \centering

    \begin{subfigure}[t]{0.48\linewidth}
        \centering
        \includegraphics[width=\linewidth]{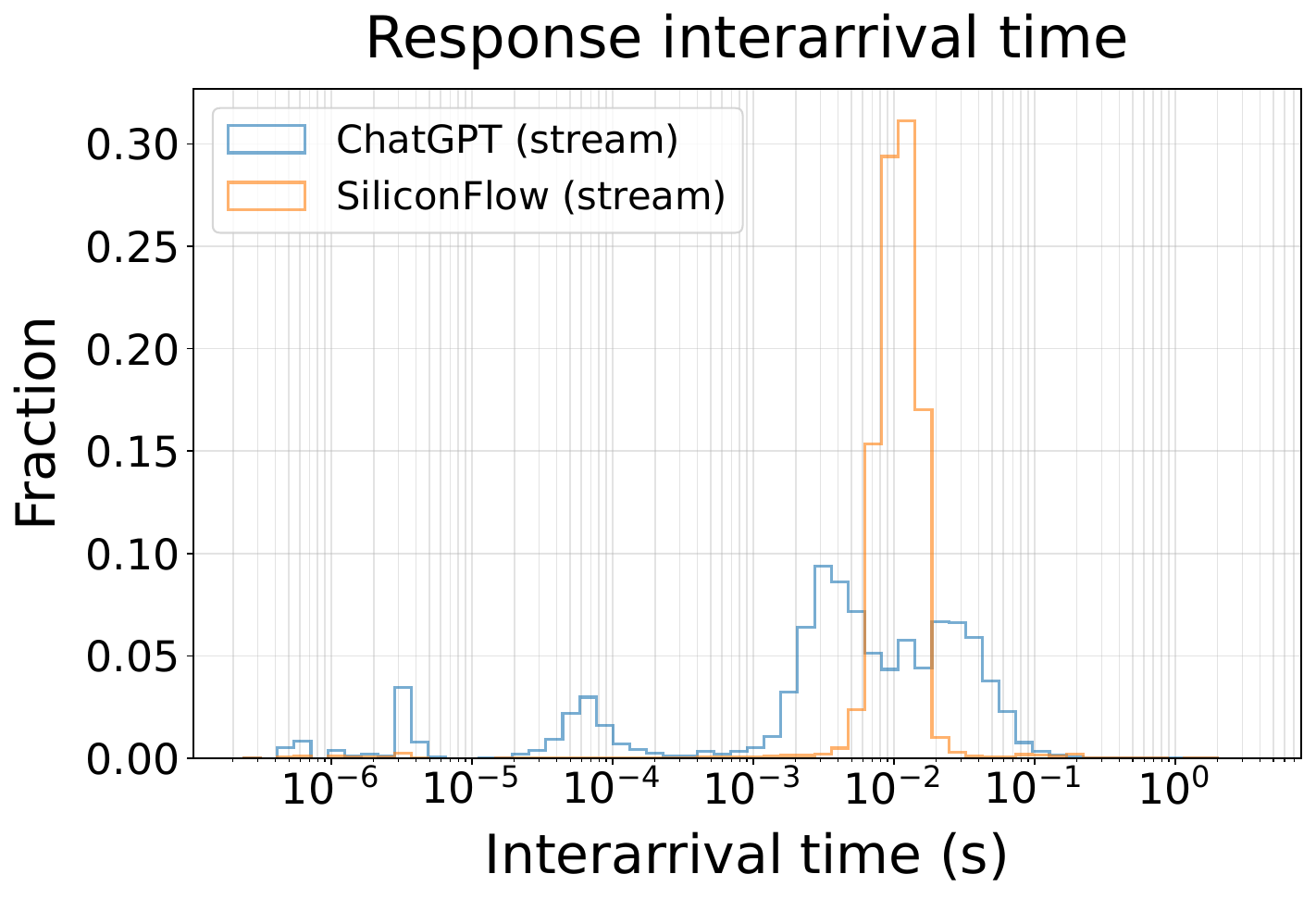}
        \label{fig:response_iat}
    \end{subfigure}
    \hfill
    \begin{subfigure}[t]{0.48\linewidth}
        \centering
        \includegraphics[width=\linewidth]{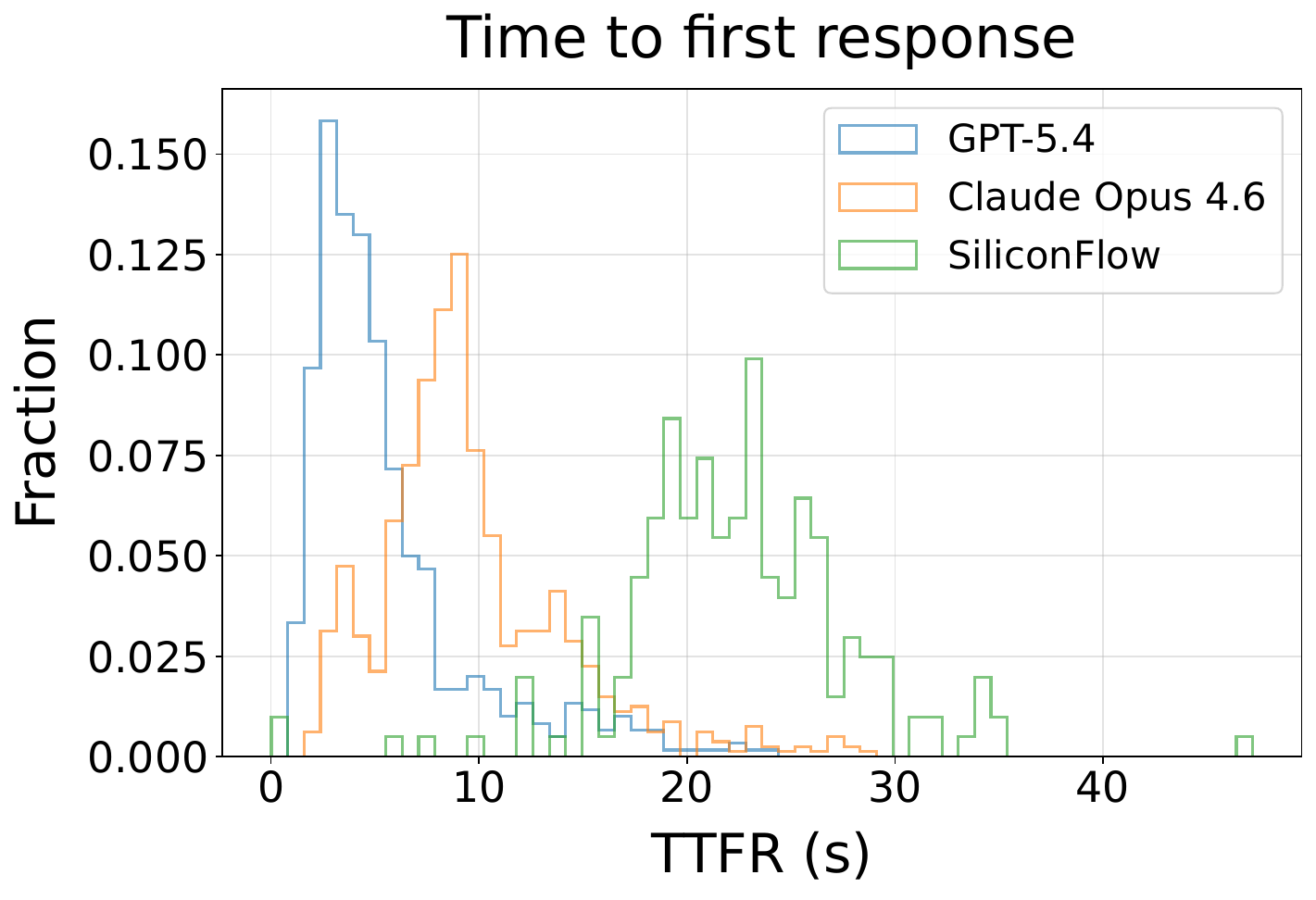}
        \label{fig:ttfr}
    \end{subfigure}

    \caption{Distribution comparison of different models in (a) T2T-Stream response interarrival time and (b) T2T-Nonstream time to first response.}
    \label{fig:model_comparison_timing}
\end{figure}

\end{document}